# Simulation of the impregnation in the porous media by the Self-organized Gradient Percolation method

A.-K. Nguyen [1], E. Blond [1], T. Sayet [1], E. de Bilbao [2], A. Batakis [3], M.-D. Duong [4]

[1] *PRISME EA 4229, Université d'Orléans, Orléans 45000, France, {anh-khoa.nguyen, eric.blond, thomas.sayet}@univ-orleans.fr*
[2] *CEMHTI (CNRS Orléans), Av. rech. scientifique, 45072 Orléans, France, emmanuel.debilbao@univ-orleans.fr*
[3] *MAPMO, UMR CNRS 6628, Université d'Orléans, Orléans 45000, France, athanasios.batakis @univ-orleans.fr*
[4] *University of Science, 227 Nguyen Van Cu, Hochiminh city, Vietnam, dmduc@hcmus.edu.vn*

**Abstract** — Many processes can correspond to reactive impregnation in porous solids. These processes are usually numerically computed by classical methods like finite element method, finite volume method, etc. The disadvantage of these methods remains in the computational time. The convergence and accuracy require a small step-time and a small mesh size, which is expensive in computational time and can induce a spurious oscillation. In order to avoid this problem, we propose a Self-organized Gradient Percolation algorithm. This method permits to reduce the CPU time drastically.

## 1. Introduction

To predict the progression of the non-reactive impregnation in the case on unsaturated sample, the classical methods [3] [4] use the Richard's equation (Eq. **(1)**) combined with Darcy's law (Eq. **(5)**). Richard's equation can be written as follows:

$$\frac{\partial(\rho\phi S)}{\partial t} = -div(\rho\vec{q}) \tag{1}$$

with

$$\phi = V_P/V_T \tag{2}$$

and

$$S = V_f/V_P \tag{3}$$

where $\rho$ is the liquid density, $\phi$ designates the porosity, $S$ designates the saturation of the pores, $\vec{q}$ is the flow density vector, $V_P$ designates the porous volume, $V_T$ designates the bulk volume and $V_f$ designates the porous volume occupied by fluid.

In the isotropic case, Darcy's law is given by:

$$\vec{q} = -k\,\underline{grad}(P_{cap}) \tag{4}$$

In the unsaturated case, the Eq. **(4)** is extended to

$$\vec{q} = -\Psi(S)\frac{K_{int}}{\eta}\underline{grad}(P_{cap}) \tag{5}$$

where $,-\underline{grad}(P_{cap})$ is the driving force, i.e., capillary force using Darcy's law, $k$ is the permeability of the porous solid, $K_{int}$ is the intrinsic permeability, $P_{cap}$ is the capillary pressure, $\eta$ is the dynamic viscosity, $\Psi(S)$ is the relative permeability, which depends on the saturation. Herein, the driving force is determined as follows:

$$-\underline{grad}(P_{cap}) = -\frac{\partial P_{cap}}{\partial S}\underline{grad}(S) \qquad (6)$$

where $\frac{\partial P_{cap}}{\partial S}$ is deduced from the capillary pressure curve. Rewriting Eq. **(5)** combined with Eq. **(6)** gives

$$\vec{q} = -\Psi(S)\frac{K_{int}}{\eta}\frac{\partial P_{cap}}{\partial S}\underline{grad}(S) \qquad (7)$$

Thus, rewriting Eq. **(1)** combined with Eq. **(7)** gives

$$\Phi\frac{\partial S}{\partial t} = div\left(\frac{K_{int}\Psi(S)}{\eta}\frac{\partial P_{cap}}{\partial S}\underline{grad}(S)\right) \qquad (8)$$

In order to obtain an optimal solution of Eq. **(8)**, it's necessary to determine the capillary pressure curve, $P_{cap}(S)$, which is a state function of the fluid saturation. There are some models to reproduce such curve in the literature. We can cite the phenomenological models [5], the extended Boltzmann transform function [8], the morphological pore network model [7], the percolation theory [6], etc. After that, the derivative of capillary pressure curve is injected in the Eq. **(8)**. Then, Eq. **(8)** can be solved by numerical methods such as the finite element method (FEM), the finite volume method (FVM), and so on [3] [4]. However, these numerical methods are expensive in computational time. Moreover, the accuracy in transient stage requires a small step time, but too low step time induces a spurious oscillation in space linked to the mesh size [9]. Hence, it's necessary to overcome these disadvantages, and more especially the high computational cost. Therefore, we propose the Self-organized Gradient Percolation model.

## 2. Problem formulation

### 2.1. Introduction to the SGP model

The aim of the Self-organized Gradient Percolation model is to predict the capillary pressure curve at any time using a proposed capillary pressure curve at initial step time (Figure 1 on the left).

The capillary pressure curve at each time step can be assimilated to a Probability Density Function (PDF) by a simple revert of the coordinate system (Figure 1 on the right). Herein, the probability density function is defined as follows:

$$S(h) = S_{max}exp\left(-\frac{h^{2m}}{m\sigma^m}\right) \qquad (9)$$

where $S_{max}$ designates the maximum saturation, $h$ designates the height, and $\sigma$ designates the variance.

In particular, if $m = 2$, then $S(h)$ will be a probability density function for a Normal distribution, and if $m = 1$, then $S(h)$ will be a probability density function for a Laplace distribution.

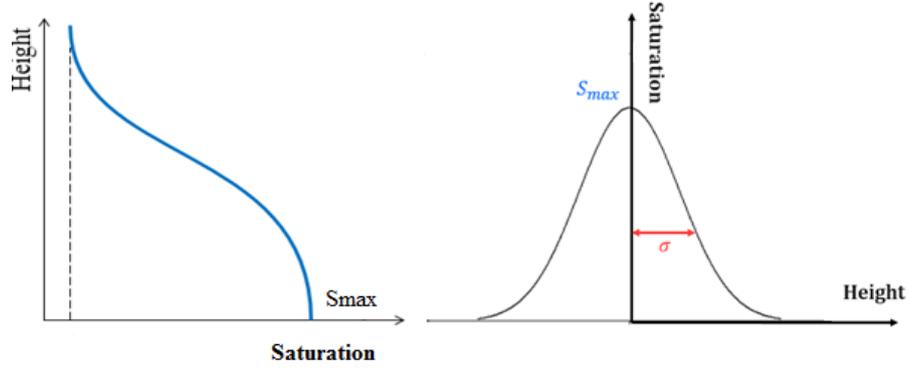

Figure 1 –On the left: principle of the sorption curve [2]; On the right: the sorption curve is assimilated to a PDF after the revert of the coordinate system.

Then, it is assumed that the equation of the capillary curve at any time step can be obtained from the previous step considering an increase of the variance. The incremental increase of the variance will be directly related to the driving force.

### 2.2. Algorithm of the SGP model

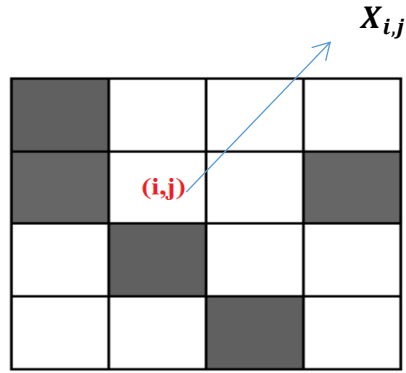

Figure 2 – The site square lattice: $X_{i,j}$ is the value at site $(i,j)$ where $i,j$ are the index of row and the index of column respectively.

In the SGP model, a porous sample is modeled as a site square lattice where $i,j$ are the index of row and the index of column respectively. A local average porosity is assign for each square. Furthermore, a random value $X_{i,j}$ from the probability density function used to model the capillary pressure curve (Figure 2) is assign to each square, depending on its position considering that $i$ related to the height. Then, the local saturation for each square is defined by:

$$S_{i,j} = X_{i,j} * f(i,j) \qquad (10)$$

where $\langle * \rangle$ is convolution operator and $f(i,j)$ is a function which makes the result of the SGP model become continuous and reflects the effects of boundary conditions. The capillary pressure curve at the initial time-step is given in the left side of the Figure 3. It is the average value of the saturation at each row in the square lattice written as:

$$\mu(0,h) = S_{max} exp\left(-\frac{|h - h_0^{S_{max}}|^m}{m[\sigma(0,h)]^m}\right) \qquad (11)$$

where $m$ is a parameter for the type of the distribution, $\mu$ is the average value of the saturation at each row in the square lattice, $h_n^{S_{max}}$ is the maximum height of the fully saturation at time step $n$, and $\sigma(n,h)$ is variance of the probability density function used to model the capillary pressure curve at a time-step $n$ and for the height $h$. The capillary pressure curve in the second time step corresponds to the initial capillary pressure curve combined with an incremental increase of the variance as follows:

$$\mu(1,h) = S_{max} exp\left(-\frac{|h - h_1^{S_{max}}|^m}{m[\sigma(0,h) + c(0,h)]^m}\right) \tag{12}$$

where $c(0,h)$ is the incremental increase of the variance. According to our assumptions (Figure 1, Eq. **(12)**), the incremental increase in variance can be written as follows:

$$c(0,h) = \left(\frac{\mu(0,h)}{\mu^*}\right)^{(m-1)/m} \left(\frac{K_{int}\rho_W g}{\eta\Phi(1 + [2ln(\mu^*/S_{max})]^{1/m})}\right) \tag{13}$$

where $\rho_W$ is the mass density of the fluid, $g$ is the gravity and $\mu^*$ is a fixed reference of $\mu$. The maximum height of the fully saturation at the second time step is

$$h_1^{S_{max}} - h_0^{S_{max}} = c(0,h)\left(\frac{\mu(0,h)}{\mu^*}\right)^{(1-m)/m} \left(1 + [2ln(\mu^*/S_{max})]^{1/m}\right) \tag{14}$$

As a result, we obtain the general capillary pressure curve in the form:

$$\mu(n,h) = S_{max} exp\left(-\frac{|h - h_n^{S_{max}}|^m}{m[\sigma(n-1,h) + c(n-1,h)]^m}\right) \tag{15}$$

where

$$c(n-1,h) = \left(\frac{\mu(n-1,h)}{\mu^*}\right)^{(m-1)/m} \left(\frac{K_{int}\rho_W g}{\eta\Phi(1 + [2ln(\mu^*/S_{max})]^{1/m})}\right) \tag{16}$$

and

$$h_n^{S_{max}} - h_{n-1}^{S_{max}} = c(n-1,h)\left(\frac{\mu(n-1,h)}{\mu^*}\right)^{(1-m)/m} \left(1 + [2ln(\mu^*/S_{max})]^{1/m}\right) \tag{17}$$

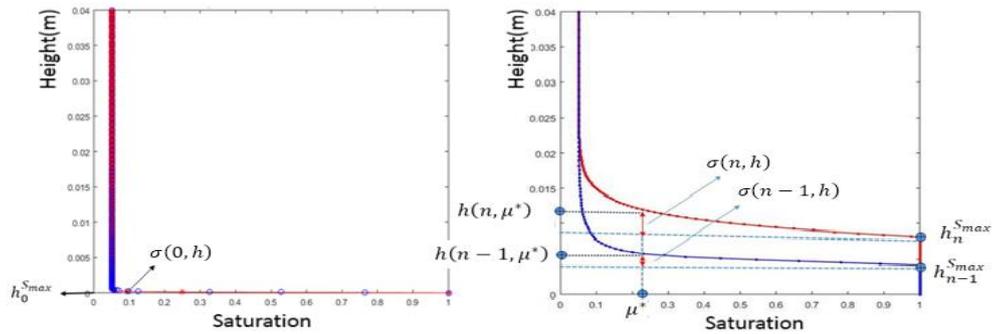

Figure 3– On the left: the capillary pressure curve at initial time step; on the right: the capillary pressure curve at general time steps

The algorithm of the SGP model is summarized in the diagram 1.

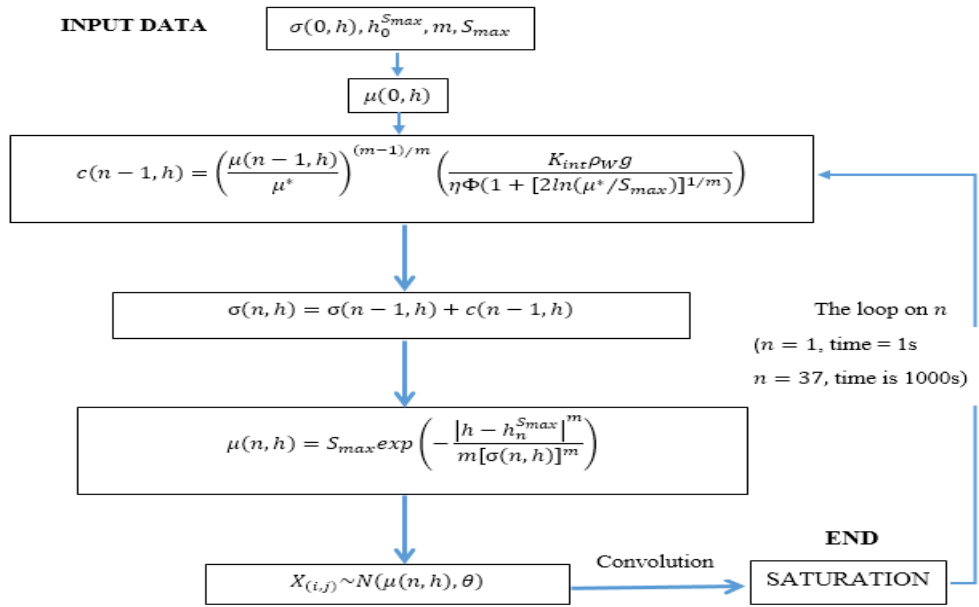

Diagram 1- The algorithm of the SGP model

## 3. Comparison with experimental results and F.E.M. results

### 3.1 Experiments description

The samples used for the experiments are cylinder with 40 $mm$ in height and 35 $mm$ in diameter. The porous sample is hung upon a free surface of liquid (Figure 4). Two different tests were done. The materials and the liquid used for each test are given in table 1. The material parameters are sum up in the table 2.

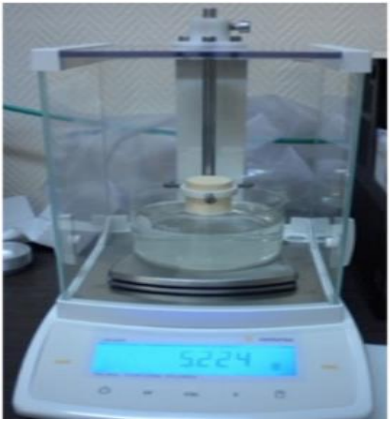

Figure 4- The non-reactive impregnation test

Table 1 – The material and liquid for the tests

| Test | Material | Liquid |
|------|----------|--------|
| 1 | Alumina 99% (AL25) | Glycerine |
| 2 | Carbon LCC | Oil |

Table 2 – The parameters used for the porous material and the liquid

| Properties | Values | | Units |
|---|---|---|---|
| | Test 1 | Test 2 | |
| Initial porosity | 0.2 | 0.19 | |
| Viscosity of the fluid, $\eta$ | 1.02 | 0.35 | $Pa \cdot s$ |
| Mass density of fluid, $\rho_W$ | 1260 | 892 | $kg \cdot m^{-3}$ |
| Intrinsic permeability, $K_{int}$ | $9.5 \times 10^{-12}$ | $7.44 \times 10^{-11}$ | $m^2$ |

In order to test the SGP model, we decide to build two FEM models using the two most used model for the capillary pressure curve [5]: Van Genuchten model, the Brooks and Corey model. Both of these models have been computed with the FE method code 'ASTER' [2].

### 3.2. Model 1 : Van Genuchten model

The capillary pressure curve according to Van Genuchten model [2] [5] is given by the following form:

$$P_{cap}(S) = P_0 \left( S^{-1/l} - 1 \right)^l \tag{18}$$

where $P_0$ is a pressure reference value and $l$ is an empirical parameter. This model allows to fit well the mass gain versus time obtained from the first experimental test.

### 3.3. Model 2: Brooks and Corey model

The capillary pressure curve according to Brooks and Corey model [5] is given by the form:

$$P_{cap}(S) = P_e S^\lambda \tag{19}$$

where $P_e$ designates the reference pressure and $\lambda$ is an empirical parameter. This model allows to fit well the mass gain versus time obtained from the second experimental test.

### 3.4. SGP model

To solve Eq. **(15)**, **(16)** and **(17)**, we fixed the initial value of the height of maximum saturation, the variance, the maximal saturation and the type of the distribution (i.e. value of $m$).

Table 3 – The parameters used for the SGP model

| Paremeters | Value for test 1 | Value for test 2 |
|---|---|---|
| $h_0^{S_{max}}$ | 0 | 0 |
| $m$ | 2 | 1 |
| $S_{max}$ | 1 | 1 |
| $\sigma(0, h)$ | $10^{-4}$ | $10^{-4}$ |

## 3.5. Comparison

The figure 4 presents the comparison between the results from the Van Genuchten model (FE Model), the SGP model and, for the mass gain, the result from SGP model and experimental results.

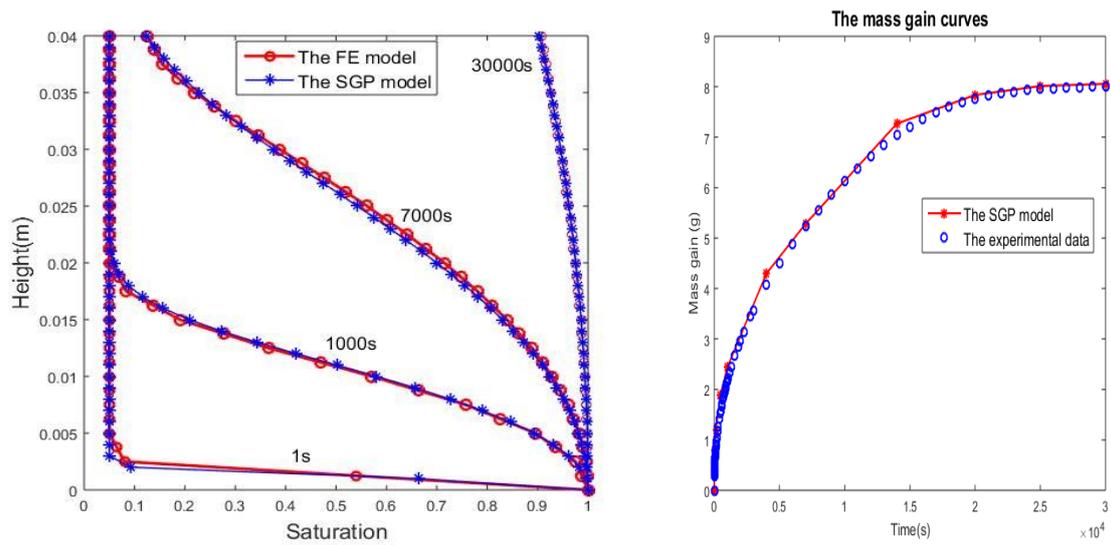

Figure 4– On the left: the evolution of the capillary pressure curve for the test 1 at different time steps; on the right: the mass gain

The figure 5 presents the comparison between the results from the Brooks and Corey model (FE Model), the SGP model and, for the mass gain, the result from SGP model and experimental results

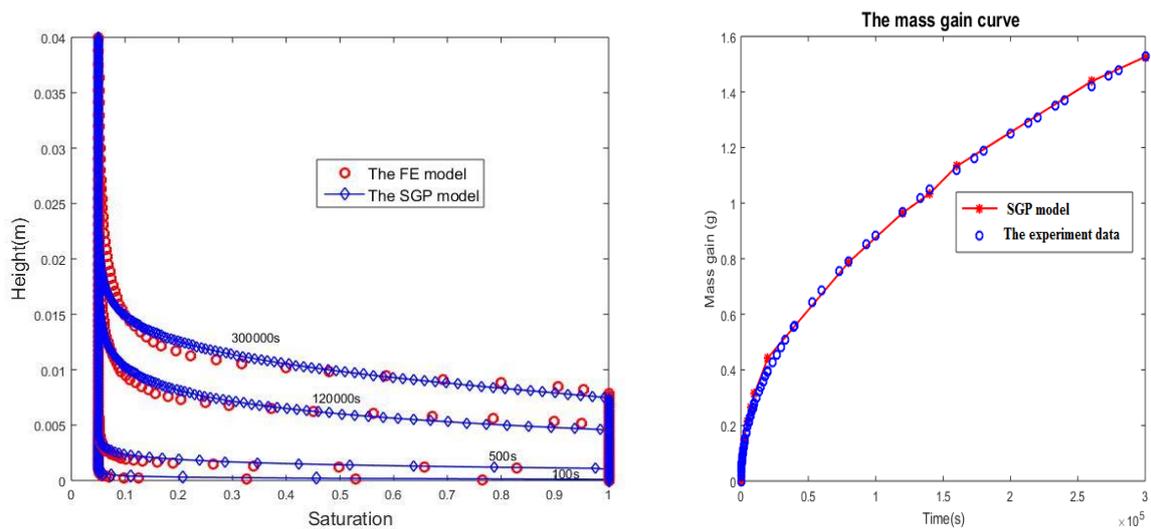

Figure 5–On the left: The evolution of the capillary pressure curve for the test 2 at different time steps; on the right: the mass gain

We obtain a very good agreement between the SGP model results and the finite element method results for the capillary pressure. We also obtain a good agreement between the SGP model results and the experiment results for the mass gain. The table 4 shows the CPU time for the different tests and models. The table 4 shows that the SGP model reduced the CPU time by a factor of **10** for the test 1, and a factor of 100 for the test 2.

Table 4 – CPU time

| Test | CPU time using FE model | CPU time using SGP model |
|---|---|---|
| 1 | 52 s | 6.517 s |
| 2 | 580 s | 1.119 s |

## 4. Conclusions

In this work, we developed a novel algorithm based on the Self-organized Gradient Percolation method. According to the value of m, we can easily modify the type of the considered capillary pressure curve, following Van Genuchten or Brooks and Corey model or other. Besides, the convolution operator makes the result of the SGP model continuous and permit to satisfies the boundary conditions.

These firsts results confirm that it is possible to reduce drastically the CPU time using the SGP method. Yet, there still being a huge work to do in order to study the mesh sensitivity, the impact of the time step, the impact of each of the three parameters of the model and their link with the physics (properties of the material).